\documentclass[
  journal=pasa,
]{cup-journal}

\usepackage{microtype,siunitx,booktabs}
\usepackage{float}
\usepackage{graphicx}	
\usepackage{amsmath}	
\usepackage{amssymb}	
\usepackage{multicol}        
\usepackage{bm}		
\usepackage{pdflscape}	
\usepackage{multirow}
\usepackage{subcaption}
\usepackage{xcolor}
\usepackage{hyperref}
\RequirePackage{natbib}
\title{Milliarcsecond Structures of Variable Peaked-Spectrum Sources}

\author{K. Ross}
\affiliation{International Centre for Radio Astronomy Research, Curtin University, Bentley, WA 6102, Australia}
\email[K. Ross]{kathryn.ross@icrar.org}

\author{C. Reynolds}
\affiliation{
CSIRO, Space and Astronomy, P.O. Box 1130, Bentley, WA 6102, Australia
}

\author{N. Seymour}
\affiliation{International Centre for Radio Astronomy Research, Curtin University, Bentley, WA 6102, Australia}

\author{J. R. Callingham}
\affiliation{Leiden Observatory, Leiden University, PO Box 9513, Leiden, 2300\,RA, The Netherlands}
\alsoaffiliation{ASTRON, Netherlands Institute for Radio Astronomy, Oude Hoogeveensedijk 4, Dwingeloo, 7991 PD, The Netherlands}

\author{N. Hurley-Walker}
\affiliation{International Centre for Radio Astronomy Research, Curtin University, Bentley, WA 6102, Australia}

\author{H. Bignall}
\affiliation{Manly Astrophysics, 15/41-42 East Esplanade, Manly, NSW 2095, Australia}
\alsoaffiliation{
CSIRO, Space and Astronomy, P.O. Box 1130, Bentley, WA 6102, Australia
}

\doi{}

\received {03 Aug 2022}
\revised  {24 Nov 2022}
\accepted {31 Dec 2022}
\published{XX}


\begin{document}

\begin{abstract}
    Spectral variability offers a new technique to identify small scale structures from scintillation, as well as determining the absorption mechanism for peaked-spectrum (PS) radio sources. In this paper, we present very long baseline interferometry (VLBI) imaging using the Long Baseline Array (LBA) of two PS sources, MRC\,0225--065 and PMN\,J0322--4820, identified as spectrally variable from observations with the Murchison Widefield Array (MWA). We compare expected milliarcsecond structures based on the detected spectral variability with direct LBA imaging. We find MRC\,0225--065 is resolved into three components, a bright core and two fainter lobes, roughly $430$\,pc projected separation. A comprehensive analysis of the magnetic field, host galaxy properties, and spectral analysis implies that MRC\,0225--065 is a young radio source with recent jet activity over the last $10^2$--$10^3$\,years. We find PMN\,J0322--4820 is unresolved on milliarcsecond scales. We conclude PMN\,J0322--4820 is a blazar with flaring activity detected in 2014 with the MWA. We use spectral variability to predict morphology and find these predictions consistent with the structures revealed by our LBA images. 
\end{abstract}

\section{Introduction}
\label{sec:intro}
Peaked-spectrum (PS) sources, are a subset of active galactic nuclei (AGN) that are identified by a peak in their radio spectral energy distribution \citep{2021A&ARv..29....3O}, and are also often associated with compact morphologies \citep[$\lesssim20$\,kpc;][]{1982A&A...106...21P,2010AJ....140.1506T}. PS sources provide an interesting population of AGN as the evolutionary pathway from PS source to extended ($\gtrsim 30$\,kpc) AGN is still unclear. Two contending theories hypothesise the nature and evolutionary pathway of PS sources: the \emph{youth scenario}, where the age of the PS source is $\leq10^5$\,years and has not yet had ample time to grow to the large-scale AGN \citep{1997AJ....113..148O,1998A&A...337...69O,2006A&A...445..889T}; and the \emph{frustration scenario}, when the PS source is confined by a dense cloud of the interstellar medium (ISM) of the host galaxy environment \citep{1984AJ.....89....5V,1984Natur.308..619W,1991ApJ...380...66O}. Furthermore, recent identifications of embedded PS cores within remnant ageing lobes has been attributed to restarted and episodic AGN activity \citep{2019MNRAS.489.4049H}, i.e. a cyclical evolution rather than linear evolution. 

Compact symmetric objects (CSOs) are a subset of PS sources with similar morphologies to large scale AGN, namely a central region (often quite faint, if detected) with emission either side associated with hot spots and/or lobes. Unlike typical AGN, CSOs show emission only on very compact scales, typically $\leq$1\,kpc, and thus require high-resolution imaging to detect \citep{1982A&A...106...21P,2005ApJ...622..136G}. CSOs are generally considered young AGN \citep[$<10^4$\,yr;][]{1997AJ....113..148O,1998A&A...337...69O,2006A&A...445..889T}, which may evolve into typical, radio-loud AGN.

Previous attempts to discriminate between youth and frustration scenarios have relied on spectral modelling and high-resolution imaging \citep[e.g.][]{2014ApJ...780..178M,2019A&A...628A..56K} using very long baseline interferometry (VLBI). The cause of absorption at low-frequencies, producing the spectral peak, has typically been attributed to synchrotron-self absorption (SSA) and/or free-free absorption (FFA) for the youth and frustration scenarios respectively \citep{2003AJ....126..723T,2015ApJ...809..168C}. Unfortunately, without sufficient sampling below the spectral turnover, the cause of absorption is often ambiguous \citep{2017ApJ...836..174C}. In rare cases, a SSA model can be ruled out if the optically thin spectral index is sufficiently steep ($\alpha\geq2.5$)\footnote{We assume a power-law relation where $ S_\nu = S_0 \nu^\alpha$, thus the sign of $\alpha$, being negative or positive, also indicates either the optically thin or thick spectral index respectively.}. 

Many PS sources have been identified as a CSOs \citep[e.g. 0108+388, 0710+439 and 2352+495;][]{1996ApJ...460..634R}. As CSOs are typically considered to be young AGN, identifying PS sources that are also CSOs could help to differentiate between the youth and frustrations scenarios. However, identifying CSOs requires high resolution (mas) observations using VLBI. Likewise, PS sources sometimes display extremely asymmetrical mas structures, likely due to an inhomogeneous surrounding environment influencing their growth \citep{2006A&A...450..959O,2019A&A...628A..56K}, compared with a fairly symmetrical morphology associated with CSOs with minor asymmetries likely coming from orientation effects \citep{2008A&A...487..885O}. VLBI can also be used to measure proper motion of hot-spots in lobes to estimate kinematic ages of $\leq3\times10^3$years \citep[][]{2003PASA...20...69P,2005ApJ...622..136G}, consistent with the theory that CSOs are young AGN. Indeed, \citet{2005ApJ...622..136G} find a majority of the CSOs with age estimates were $\leq500$\,yrs, suggesting CSOs may be short lived and few would continue to grow to the scale of typical AGN, thereby explaining the large fraction of CSO and PS sources thought to be young relative to the number of large-scale radio galaxies \citep[][]{2021A&ARv..29....3O}. VLBI of PS sources can thus help to identify populations of CSOs and elucidate the youth scenario and AGN evolution. 

Spectral variability at radio frequencies offers a new technique for identifying young or frustrated candidates. Many variability surveys have identified PS sources that lost their PS classification over time \citep[][hereafter R21]{2005A&A...432...31T,2005A&A...435..839T,2021MNRAS.501.6139R}, or showed a significant change in spectral shape likely due to a variable opacity from the inhomogeneous surrounding ISM \citep[][hereafter R22]{2015AJ....149...74T,2022MNRAS.512.5358R}. Thus the population of known PS sources, which is already biased from sparse spectral coverage from a range of instruments and times, is likely contaminated by temporary PS sources. This is particularly true at higher frequencies ($\sim$GHz), which is sensitive to emission from the core/jets. PS sources with a peak at lower frequencies ($\sim$MHz) appear to be less contaminated by sources only showing a temporary peak \citep[][R21]{2017ApJ...836..174C}.

Spectral variability offers the a new technique to find and exclude contaminating ``temporary'' PS sources, as well as identify CSO candidates with a decreased risk of contaminating sources. Variability of PS sources has been used to infer the presence of compact ($\mu$as -- mas) features based on scintillation \citep[][R21]{Fanti_1979,ipsII}. Such compact features are common for CSOs, but VLBI is required for confirmation of a CSO classification. Spectral variability has also found PS sources that show changing spectral shape, inconsistent with scintillation, which suggests that some PS sources are frustrated or contaminating blazars (R22). 

This paper aims to investigate the milliarcsecond scale structures of variable PS sources using VLBI to test predictions based on spectral variability. In particular, we investigate PS sources that have shown a consistent spectral shape with a variable overall flux density, consistent with scintillation, suggesting a compact feature on milliarcsecond scales (R21, R22), and use VLBI to test a CSO classification. We also investigate variable PS sources that R21 found as changing spectral shape. They concluded the short timescale ($\sim$1\,year), and variable spectral shape is inconsistent with interstellar scintillation and present it as a blazar caught flaring. 

In Section~\ref{sec:target_selection}, we describe the three variable PS sources of this study, in Section~\ref{sec:methods} we describe the observational strategy and data reduction. Section~\ref{sec:results} outlines the results of the LBA imaging. We discuss the host galaxy properties including their linear size compared to turnover in Section~\ref{subsec:ls_vs_t}, the mid-infrared (MIR) and optical emission in Section~\ref{subsec:hostgals} and the radio properties in Section~\ref{subsec:radioproperties}. In Section~\ref{sec:unified_perspective} we present the likely absorption mechanisms and source classification of our targets. We adopt the standard $\Lambda$-cold dark matter cosmological model, with $\Omega_{\rm M} = 0.286$, $\Omega_\Lambda = 0.714$, and the Hubble constant $H_0 = 69.6$\,km\,s$^{-1}$\,Mpc$^{-1}$ \citep{2006PASP..118.1711W,lamdacdm}

\section{Target Selection}
\label{sec:target_selection}
Targets were selected for LBA imaging with the goal of comparing direct imaging of milliarcsecond structures with predicted morphologies based on their variability. Three targets were selected based on the variability detected by R21 and R22. MRC\,0225--065 (GLEAM\,J022744-062106) was initially identified as variable in R21 but further monitoring over a year found no evidence of variability (R22). As such, it was predicted MRC\,0225--065 would have resolved structures on milliarcsecond scales with a compact feature $\lesssim25$\,mas, resulting in variability from refractive interstellar scintillation (RISS) on a longer timescale with a dampened modulation index due to the extended structure. Conversely, PMN\,J0322--4820 (GLEAM\,J032237--482010) was selected due to the variable spectral shape identified in R21. To explain the variable spectral shape, R21 concluded PMN\,J0322--4820 was likely a blazar caught flaring in 2014. As such, it was predicted to show a compact morphology even on milliarcsecond scales. Finally, MRC\,2236-454 (GLEAM\,J223933--451414) was identified by R21 as the only PS source in their sample that showed significant variability but maintained a constant peak frequency below 231\,MHz. A low peak frequency is typically associated with PS sources that are of the order of tens of kilo-parsecs across, but the RISS detected by R22 suggested MRC\,2236-454 is dominated by a compact feature, and showed variability due to a surrounding inhomogeneous environment. As such, it was predicted MRC\,2236-454 may be resolved on milliarcsecond scales and show an asymmetrical morphology, often associated with frustrated sources in an inhomogeneous surrounding environment \citep{2006A&A...450..959O}. 

\section{LBA Observations and Data Reduction}
\label{sec:methods}

\subsection{Observations}
\label{subsec:methods_obs}
LBA observations were taken on November 23, 2020 and February 17, 2021 as part of project V600. The November observation was centered at 2.4\,GHz and the February observation was centered at 8.3\,GHz and both utilised 128\,MHz of bandwidth in dual polarizations. Stations used in each observation and their diameter is listed in Table~\ref{tab:LBA_stations}. Both observations cycled through phase calibrator scans and target scans of lengths 2\,min and 5\,min, respectively. However, the spatial separation of each target and their respective phase calibrator meant each target had a different number of scans. A summary of the targets, phase calibrators and number of scans each is presented in Table~\ref{tab:tar_obsdetails}.

Parkes at 2.4\,GHz, and Katherine at both frequencies, observed using their native linear feeds. These were converted to a circular polarization basis post-correlation using the PolConvert software \citep{martividal2016}

\begin{table}[ht]
\centering
\begin{tabular}{c c c c c}
\hline
\textbf{Name} & \textbf{Code} & \textbf{Diameter (m)} & \textbf{Nov20} & \textbf{Feb21} \\
\hline
ATCA, phased up & At & 5$\times$22 & Y & Y  \\
Mopra & Mp & 22 & Y & Y \\
Parkes & Pa & 64 & Y & Y \\
Hobart & Ho & 26 & Y & Y \\
Ceduna & Cd & 30 & Y & Y \\
Yarragadee & Yg & 12 & Y & Y \\
Warkworth & Ww & 12 & Y & Y \\
Hartebeesthoek & Hh & 26 & Y & Y \\
Katherine & Ke & 12 & Y & Y \\
Tidbinbilla & Td & 34 & Y & N\\
\hline
\end{tabular}%
\caption[LBA Stations]{LBA stations included in observations} 
\label{tab:LBA_stations}
\end{table}

\begin{table}[ht]
\centering
\begin{tabular}{c c c}
\hline
\textbf{Source Name} & \textbf{Expected $S_{\mathrm{5GHz}}$ (mJy)} & \textbf{Number of scans} \\
\hline
MRC\,0225--065 & 0.238 & 27\\
PKS\,J0217+0144 (C) & 0.666  & 27 \\
PMN\,J0322--4820 & 0.112 & 40 \\
PMN\,J0335-4837 (C) & 0.112 & 40 \\
MRC\,2236--454 & 0.420 & 48 \\
QSO\,B2227--445 (C) & 0.386 & 48 \\
\hline
\end{tabular}%
\caption[Target Observations]{Targets, associated calibrators and number of LBA scans for each target source. } 
\label{tab:tar_obsdetails}
\end{table}

\subsection{Data Processing and Calibration}
\label{sec:dataprocessing}
After correlation, data calibration and processing were done using the NRAO's Astronomical Imaging Processing System (AIPS) \citep{Wells1985}. The calibration and flagging followed the general procedure outlined in the AIPS cookbook\footnote{The AIPS cookbook can be found here \url{http://www.aips.nrao.edu/cook.html}} and was implemented in a semi-automated script with the ParselTongue interface \citep{kettenis2006}. Initial flagging of edge channels and RFI was done using \textsc{UVFLG}. Auto-correlations were scaled to unity across the band using \textsc{ACCOR} before removing gross residual instrumental delays using \textsc{FRING} on a short scan of a bright calibrator. Complex bandpass corrections were derived using \textsc{BPASS}. The system temperature and gain calibration were applied using \textsc{APCAL}. Delay, rate and phase calibrations were determined from fringe fitting using \textsc{FRING} from each target's respective phase calibrator. A phase referenced image was created for all targets except for MRC\,0225--065, as a first pass detection of the targets to determine if a phase shift was needed. Lastly, \textsc{UVFIX} was used to apply a phase shift to the data for any sources that were $\sim$arcsecond away from the phase centre used in correlation. MRC\,0225--065 had accurate VLBI coordinates and thus did not require a phase shift. The calibrated and phase shifted data were exported to be imaged using \textsc{casa}. 

\subsection{Imaging and Self-Calibration}
\label{subsec:methods_imag}
Initial Stokes-I images were made with a quasi-natural weighting with robust parameter set to +1 \citep{1995PhDT.......238B} using the \texttt{tclean} function in \textsc{casa} \citep{2007ASPC..376..127M}. Clean boxes were used but were tightly restricted for the models used for self-calibration to avoid inducing artificial structure from the complex point-spread-function. For each image, phase only self calibration was performed and applied using the \texttt{gaincal} and \texttt{applycal} functions respectively. Due to the sparse $(u,v)$-coverage and low signal-to-noise (SNR), calibration solutions were inspected and applied without flagging solutions that had insufficient SNR. The slow rate of improvement necessitated several ($\sim$9) rounds of self-calibration. The SNR of the main component and the root-mean-squared (rms) noise of the image were inspected after each self calibration iteration to ensure each round improved the overall image quality. For each source the initial model assumed for the self-calibration was an unresolved point source to avoid inducing any morphological features. Any resolved components were included in subsequent rounds of imaging clean components and kept in the model for self-calibration if this reduced the rms noise of the image. The initial solution interval for the self calibration was set to the scan length and decreased in further rounds of self calibration. Phase only self calibration rounds were continued until the rms noise of the image increased. A final round of both phase and amplitude self calibration was then performed (provided it reduced the rms of the final image) with the solution interval set to the scan length. For MRC\,0225--065, an amplitude self-calibration was applied to both frequencies, but no amplitude self-calibration was applied to the 2.4\,GHz image of PMN\,J0322--4820.

\section{Results}
\label{sec:results}
Images of MRC\,0225--065 at both 2.4 and 8.3\,GHz are presented in Figure~\ref{fig:j0227_LBA}, and an image of PMN\,J0322--4820 at 2.4\,GHz, presented in Figure~\ref{fig:j0322-482}. Unfortunately, due to large phase errors from a pointing offset, we were unable to recover images for MRC\,2236--454 at either frequency, or for PMN\,J0322--4820 at 8.3\,GHz, this was because the source positions were beyond the observed correlated field of view for recovery in each case. For MRC\,2236--454, the pointing offset was over 11\,arcseconds for both the 2.4\,GHz and 8.3\,GHz observations, thus the phase errors from this pointing offset was beyond recovery. PMN\,J0322--4820 also had a pointing offset of $\approx11.5$\,arcseconds, however, given it was bright ($\sim0.2$\,Jy), there was sufficient sensitivity using a subset of antennas (flagging the Hartebeesthoek antenna), and a phase shift combined with self calibration to recover and image at 2.4\,GHz. However, this method was not possible at 8.3\,GHz due to the smaller field-of-view and decreased sensitivity. Henceforth, we will only discuss the results for MRC\,0225--065 and PMN\,J0322--4820. 

\begin{table}[htbp]
\centering
\begin{tabular}{c c c c c}
\hline
\textbf{Source, $\nu$ (GHz)} & \textbf{rms (mJy/beam)} & \textbf{$\theta_\mathrm{beam,maj}$} & \textbf{$\theta_\mathrm{beam,min}$} & \textbf{PA}\\
\hline
MRC\,0225--065, 2.4 & 2.7 & 9.5 & 3.2 & 7.0 \\
MRC\,0225--065, 8.3 & 1.0 & 4.4 & 2.7 & 83 \\
PMN\,J0322--4820, 2.4 & 1.0 & 30 & 17 & -54 \\
\hline
\end{tabular}%
\caption[LBA Image Properties]{Properties for each LBA image: synthesised beam size and rms background noise. } 
\label{tab:lba_image_properties}
\end{table}

\subsection{MRC B0225--065}
\label{sec:images_j0227-0621}
\begin{figure*}[htpb]
     \centering
     \begin{subfigure}[b]{0.47\textwidth}
         \centering
        \includegraphics[width=0.9\textwidth]{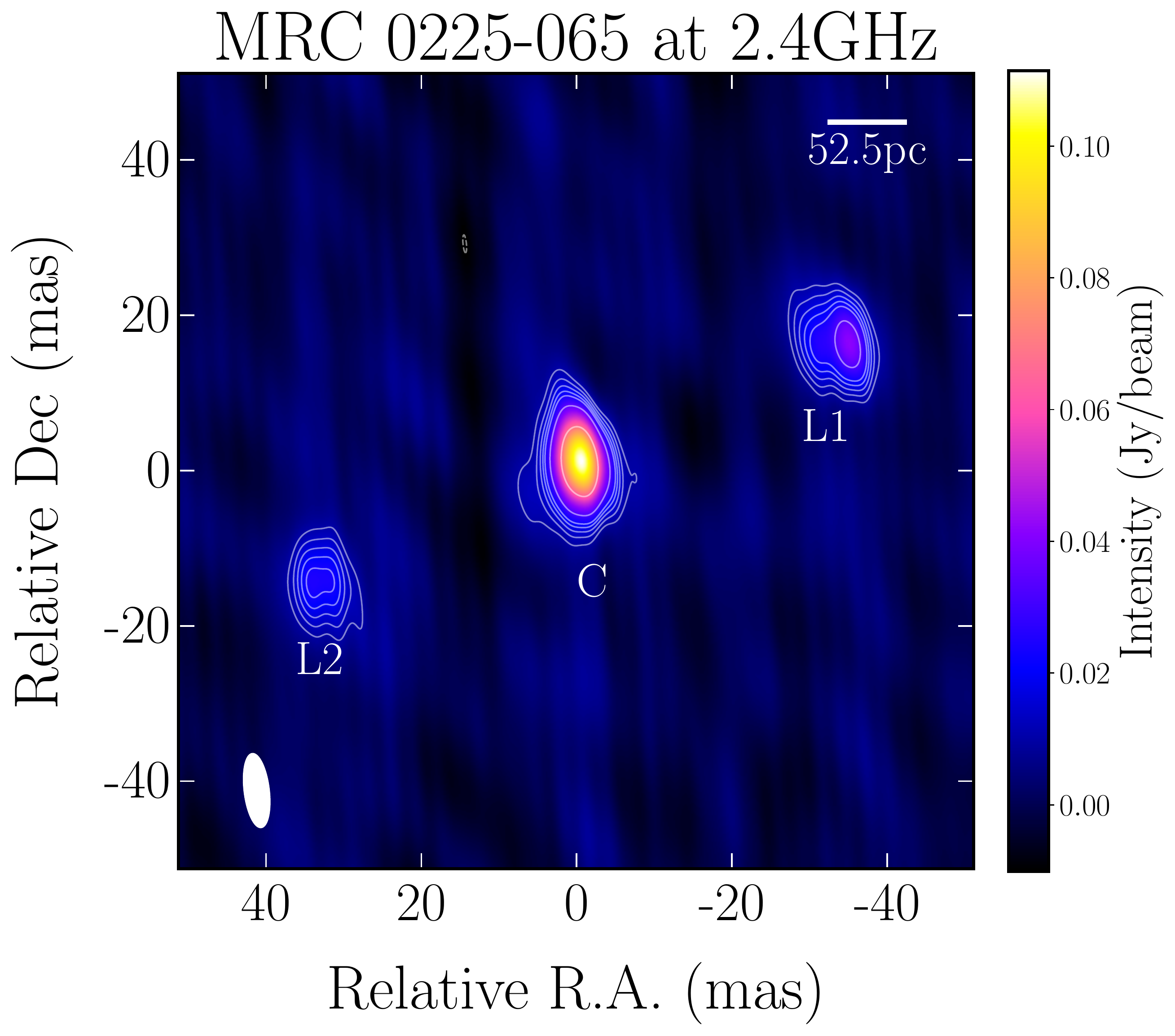}
         \label{fig:j022744_lba_2ghz}
     \end{subfigure}
     \begin{subfigure}[b]{0.47\textwidth}
         \centering
        \includegraphics[width=0.92\textwidth]{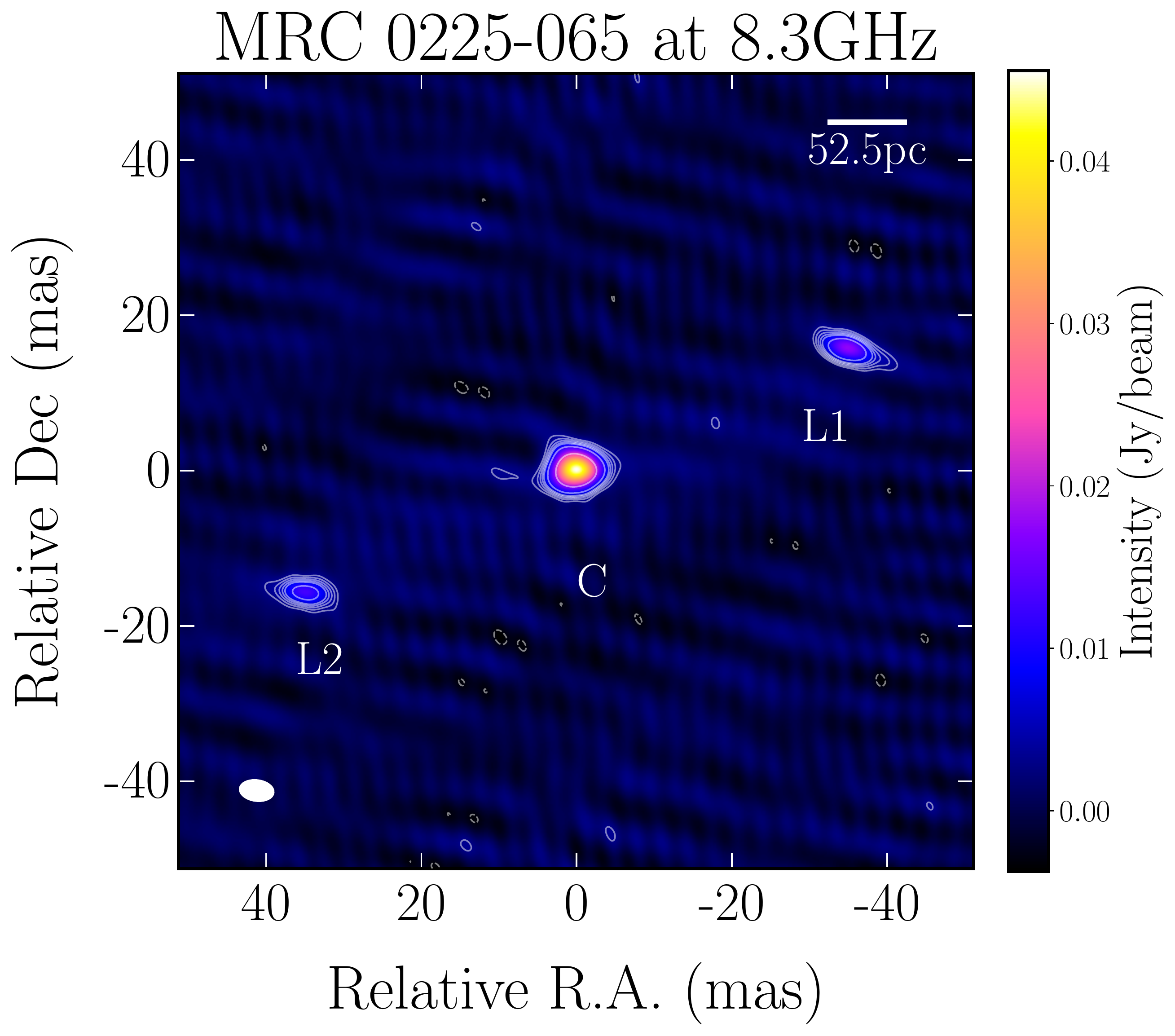}
        \label{fig:j022744_lba_8ghz}
     \end{subfigure}
     \caption[LBA Image and SED of MRC\,0225--065]{LBA images of MRC\,0225--065 at 2.4\,GHz (left) and 8.3\,GHz (right). Beam sizes are shown with a white ellipse in the bottom left corner of each image and dimensions are specified in Table~\ref{tab:lba_image_properties}. Contours are placed at (-3, 3, 4, 5, 6, 7, 10, 20, 50, 100, 200, 400, 800, 1600) times the rms noise of the image, also specified in Table~\ref{tab:lba_image_properties}. Pixel brightness is plotted in a linear scale following the colour-bars to the right of each image. The resolved regions are labelled C, L1, L2 and properties of each region are outlined in Table~\ref{tab:j0227-0621_lba_components}. Relative R.A and Dec are calculated from the position of the core (C) component with coordinates: J2000 02h27m44.5s -06d21m06.7s. }
     \label{fig:j0227_LBA}
\end{figure*}

MRC\,0225--065 was resolved into three components morphology at both 2.4\,GHz and 8.3\,GHz, as shown in Figure~\ref{fig:j0227_LBA}. The final image was made with a robust parameter of -1 at 2.4\,GHz and -0.5 at  8.3\,GHz \citep{1995PhDT.......238B}. MRC\,0225--065 is resolved into 3 regions: a bright, unresolved central component, with an upper limit of source size of $2.5\times4$\,mas assuming the beam size at 8.3\,GHz (labelled C in Figure~\ref{fig:j0227_LBA}), a fainter $16\times11$\,mas Western region (L1) and even fainter $14\times10$\,mas Eastern component (L2). The sizes of L1 and L2 are measured using the contours in the 2.4\,GHz image. The triple morphology is roughly symmetrical with the distance between the C to L1 and L2 being $\sim40$\,mas each. Since it appears the components of MRC\,0225--065 may be resolved, we measured their flux density over an irregular polygon\footnote{using \url{https://github.com/nhurleywalker/polygon-flux}, \citep{2019PASA...36...48H}} for each component.

We recovered all the flux density predictions from the spectral fit to the R22 ATCA observations at 2.4\,GHz, but found that  $\sim35$\% of the flux density was lost at 8.3\,GHz. The flux densities for each component and their spectral index are presented in Table~\ref{tab:j0227-0621_lba_components}. The irregular polygon was shaped based on contour levels to ensure only real flux was included in the final measurement. However, the missing flux density at 8.3\,GHz may be due to extended structure being resolved out. Consequently, the estimates for the spectral index presented in Table~\ref{tab:j0227-0621_lba_components} should be considered lower limits.  

\begin{table}[ht]
\centering
\begin{tabular}{c c c c}
\hline
\textbf{Component} & \textbf{$S_{\mathrm{2.4GHz}}$ (mJy)} & \textbf{$S_{\mathrm{8.3GHz}}$ (mJy)} & \textbf{$\alpha$} \\
\hline
C & 270$\pm$10 & 78$\pm$7 & -0.95$\pm$0.08 \\
L1 & 121$\pm$8 & 30$\pm$5 & -1.1$\pm$0.2 \\
L2 & 56$\pm$7 & 18$\pm$4 & -0.9$\pm$0.2 \\
Integrated LBA & 447$\pm$14 & 126$\pm$10 & -0.97$\pm$0.07 \\
Model Prediction & 400 & 195 & N/A \\
\hline
\end{tabular}%
\caption[MRC\,0225--065 Components]{Flux densities and two component spectral index for each component of MRC\,0225--065 found in the LBA images. The uncertainties for the flux densities are measured calculated using the measured uncertainty from polygon flux and the rms noise of the image. The uncertainty for $\alpha$ is calculated using standard propagation of errors. The model prediction is calculated from the best spectral fit, a double SSA spectral model with an exponential break.}  
\label{tab:j0227-0621_lba_components}
\end{table}

\begin{figure*}[htpb]
     \centering
     \begin{subfigure}[b]{0.55\textwidth}
    \includegraphics[width=\textwidth]{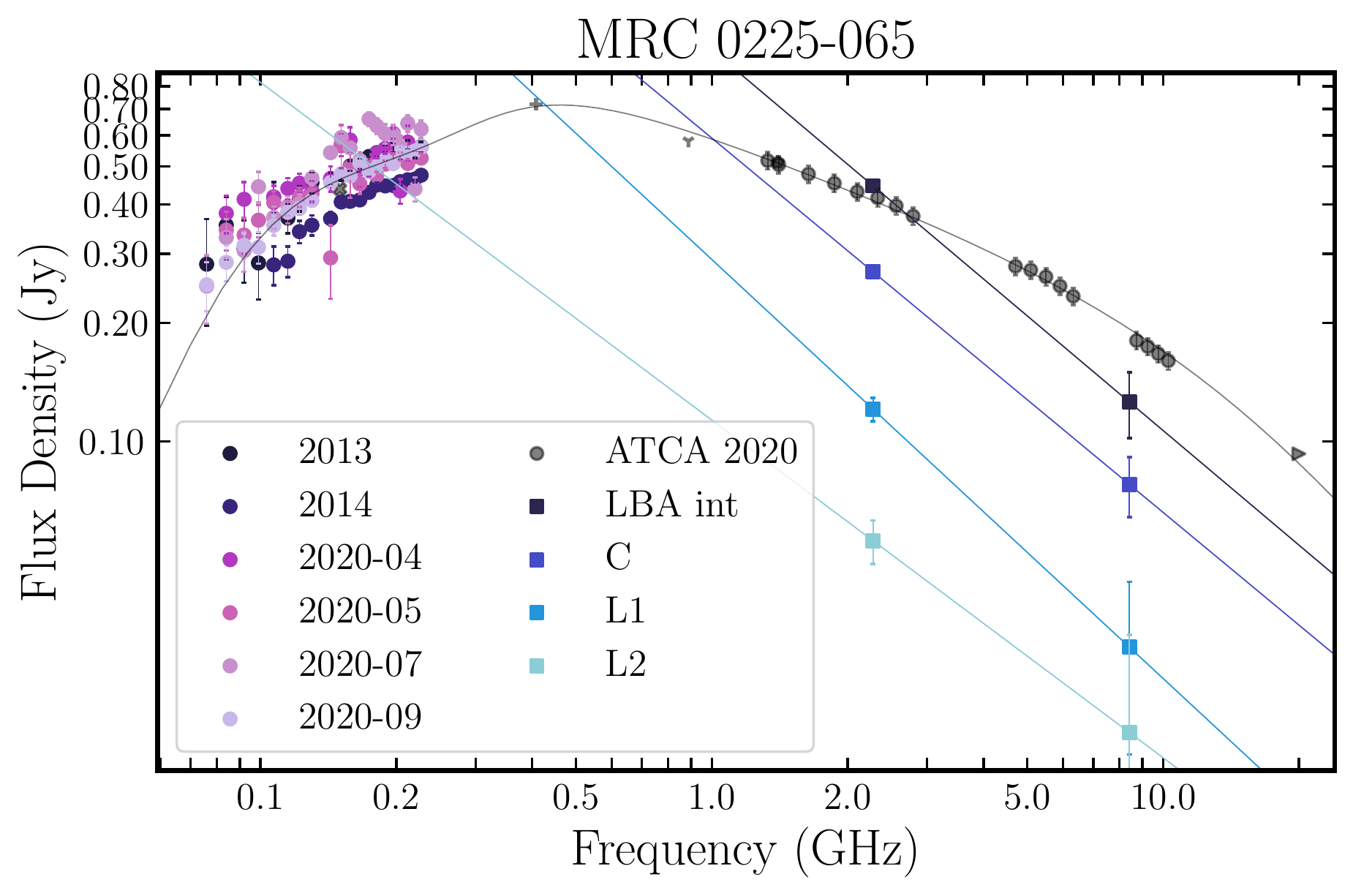}
    \end{subfigure}
     \begin{subfigure}[b]{0.42\textwidth}
    \includegraphics[width=\textwidth]{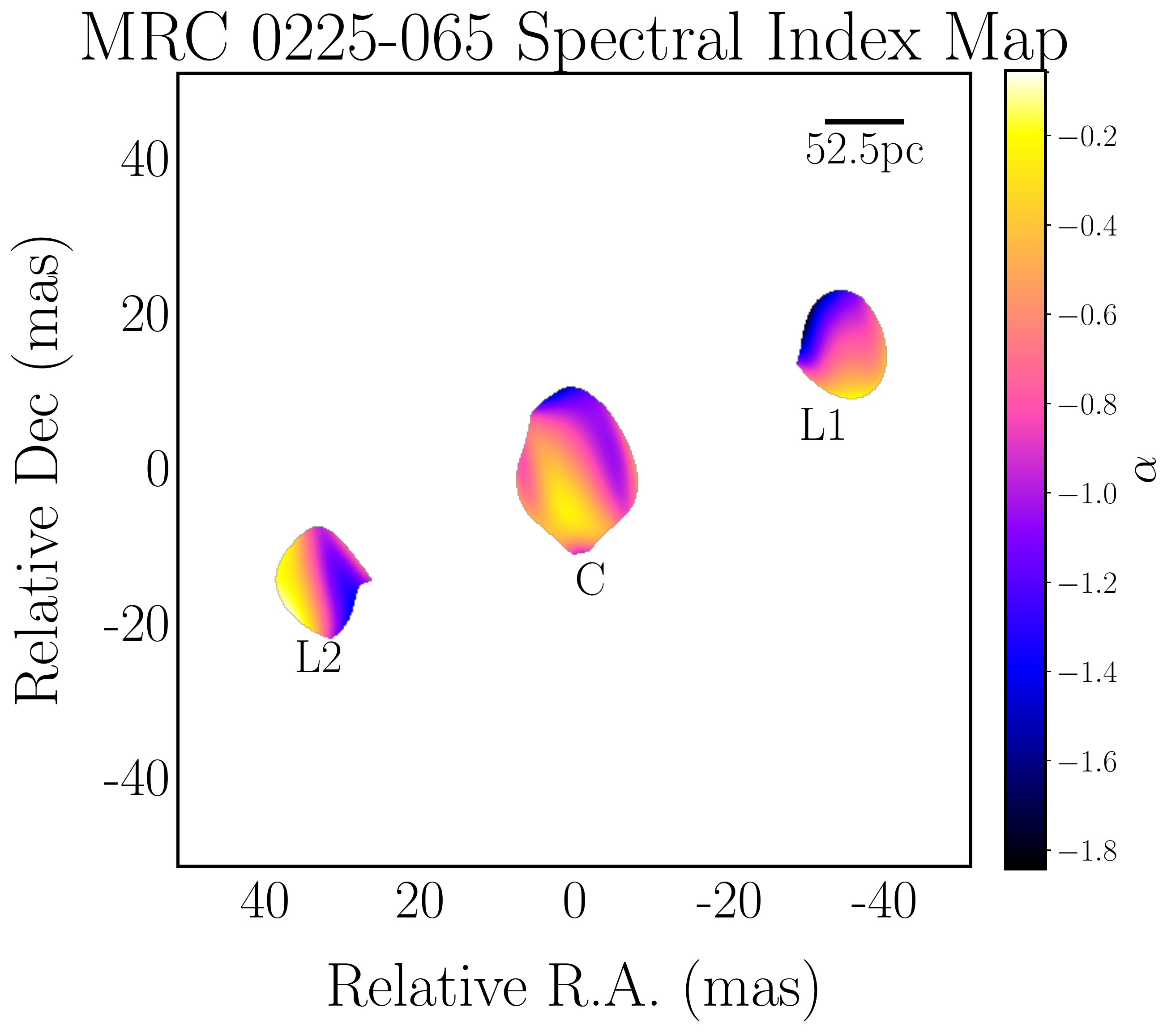} 
    \end{subfigure}
    \caption[SED for GLEAM\,J022744-062107]{Spectral energy distribution (SED) for MRC\,0225--065 (left) and spectral index map (right). The spectral index map was created using by convolving both the 8.3\,GHz image and 2.4\,GHz image to the same resolution. Data included in the SED are from R21 and R22 monitoring (circles) and coloured according to epoch. LBA flux densities are plotted as squares with the integrated flux density of LBA plotted as black squares. The spectral fit to each LBA point is a power-law with spectral index presented in Table~\ref{tab:j0227-0621_lba_components}. The grey spectral model to the entire SED is a double SSA model with an exponential break. Supplementary data included: TIFR GMRT 150\,MHz Sky Survey Alternative Data Release 1 \citep[TGSS-ADR1; ][]{intema2016TGSS} (grey cross), Molonglo Reference Catalogue \citep[MRC; ][]{1981MNRAS.194..693L,1991Obs...111...72L} (grey $+$), Rapid ASKAP Continuum Survey \citep[RACS; ][]{2020PASA...37...48M,2021PASA...38...58H} (grey `Y'), NRAO VLA Sky Survey \citep[NVSS; ][]{condon+98}, Australia Telescope 20\,GHz \citep[AT20G;][]{at20g} (grey right arrow). }
     \label{fig:j0227-0621_sed}
\end{figure*}

The symmetrical triple morphology suggests MRC\,0225--065 is a CSO candidate with a core (C) and two lobes (L1 and L2). The spectral index of the central component is $\alpha_{\mathrm{C}}=-0.95\pm0.08$, which is far steeper than expected for a typical AGN ``core", generally expected to have a $\alpha \geq-0.5$ \citep{2006A&A...450..959O,hardcastle2008properties}. However, components have previously been identified as cores with spectral indices as steep as $-0.7$ \citep{2006A&A...450..959O}. We present the SED for MRC\,0225--065 in Figure~\ref{fig:j0227-0621_sed} including the MWA flux densities from R22 as well as the flux densities and power-law spectral model for each LBA component. The entire SED is fit, using the most recent MWA epoch (2020-09), with a double SSA model with an exponential break, which assumes two synchrotron emitting regions that are self-absorbed and ageing producing the exponential break, $\nu_b$, separate from the peak frequency. The break frequency is the frequency where the spectrum begins to steepen as the electrons are ageing and experiencing energy losses \citep{2018MNRAS.474.3361T}. We fit the spectral model using the \verb|UltraNest| package\footnote{\url{https://johannesbuchner.github.io/UltraNest/}} \citep{2021JOSS....6.3001B}, which uses a nested sampling Monte Carlo algorithm. From the double SSA spectral model, we find the peak frequencies for the two SSA components to be $\nu_{p,1}=$400$\pm$100\,MHz and $\nu_{p,2}$=112$\pm$90\,MHz, and find $\nu_b=$14.3$\pm$2.7\,GHz.  

MRC\,0225--065 has a spectroscopic redshift of 0.445 \citep{2017ApJS..233...25A}; thus, 1\,mas corresponds to a linear scale of 5.25\,pc. Using this redshift, we find the projected linear size of MRC\,0225--065 (from L1 to L2) to be $\sim$430\,pc, the linear distance from the core to either lobe to be $\sim$210\,pc and place an upper limit on the size of component C to be $\leq$26\,pc.

\subsection{PMN J0322--4820}
\label{sec:images_j0322-482}
\begin{figure*}[htpb]
     \centering
     \begin{subfigure}[b]{0.49\textwidth}
         \centering
         \includegraphics[width=\textwidth]{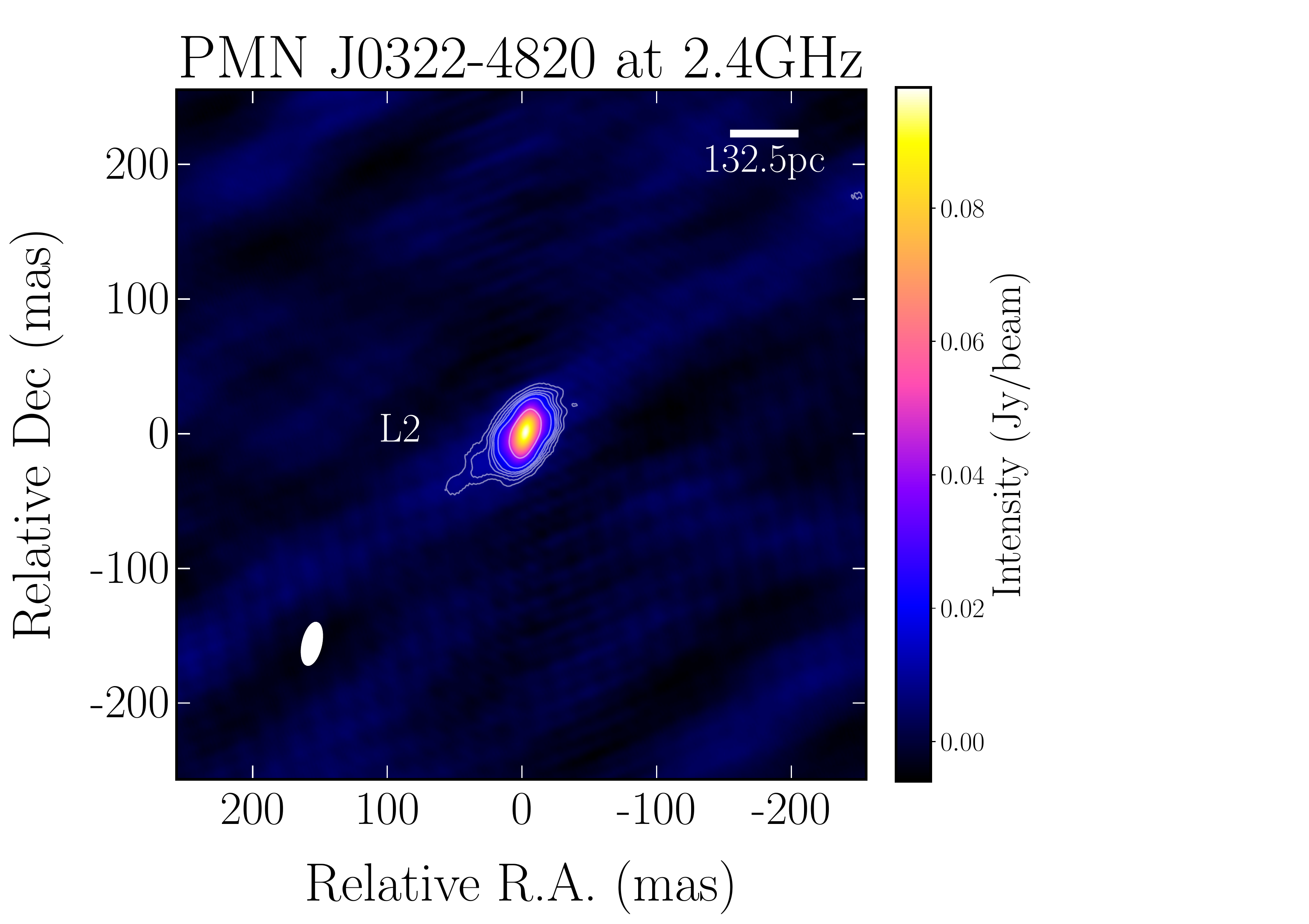}
     \end{subfigure}
     \begin{subfigure}[b]{0.49\textwidth}
         \centering
         \includegraphics[width=\textwidth]{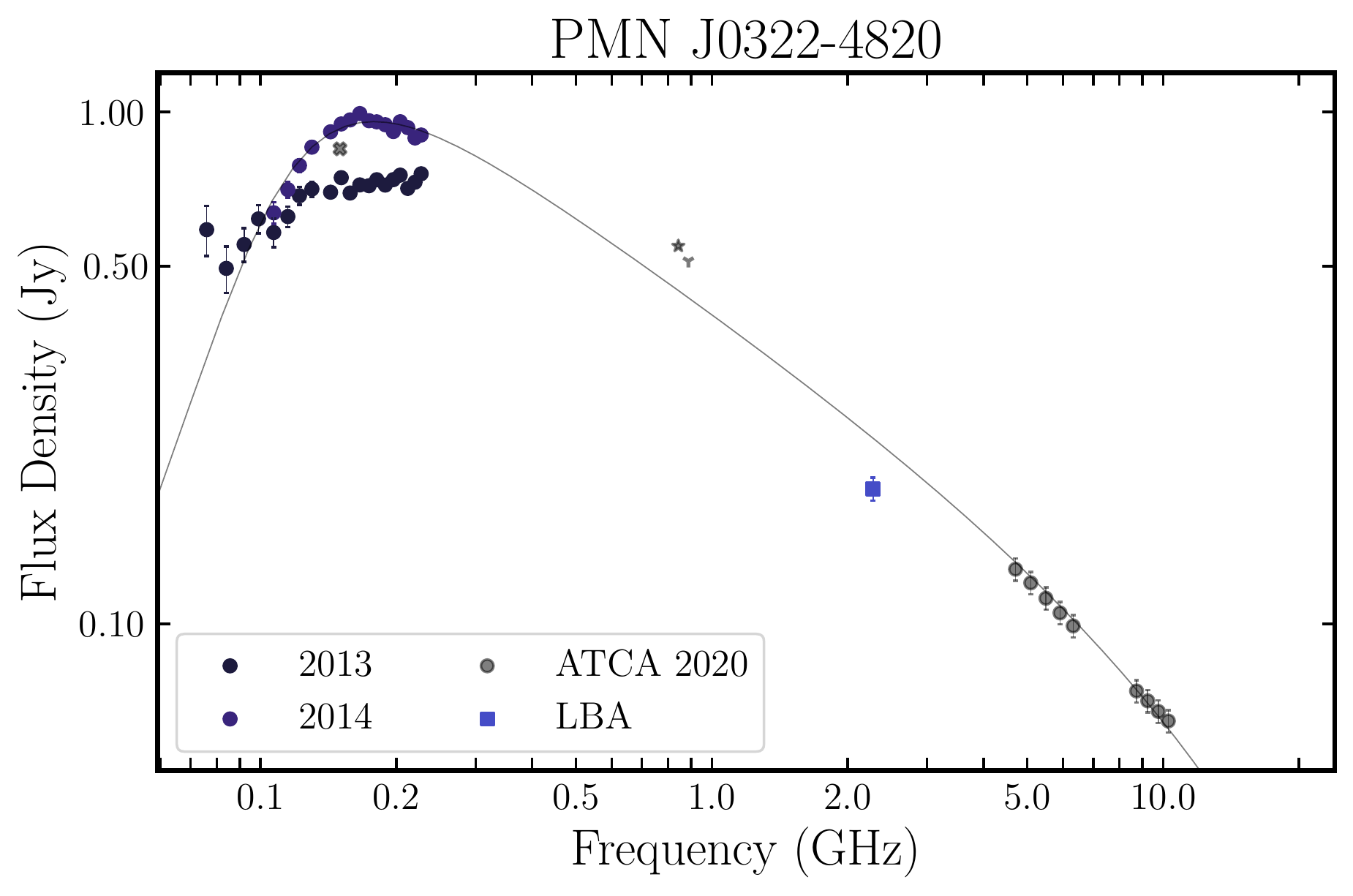}
     \end{subfigure}
     \caption[LBA Image and SED of PMN\,J0322--4820]{LBA image for PMN\,J0322--4820 at 2.4\,GHz (left) and associated SED (right). The beam size is shown with a white ellipse in the bottom left corner and dimensions are specified in Table~\ref{tab:lba_image_properties}. Contours are placed at (-3, 3, 4, 5, 6, 7, 10, 20, 50, 100, 200, 400, 800, 1600) times the rms noise of the image, also specified in Table~\ref{tab:lba_image_properties}. Pixel brightness is plotted in a linear scale following the colour-bars to the right of the image. Relative R.A and Dec are calculated from the central coordinate: J2000 03h22m38.0s -48d20m16.2s. Data included in SED is from R21 and R22 (circles) and coloured according to epoch. LBA flux density is plotted as a blue square. The grey spectral model to the entire SED is a single SSA model with an exponential break. Supplementary data included is: TIFR GMRT 150\,MHz Sky Survey Alternative Data Release 1 \citep[TGSS-ADR1; ][]{intema2016TGSS} (grey cross), Sydney University Molonglo Sky Survey \citep[SUMSS;][]{2003MNRAS.342.1117M} (grey star), Rapid ASKAP Continuum Survey \citep[RACS; ][]{2020PASA...37...48M,2021PASA...38...58H} (grey `Y'). }
     \label{fig:j0322-482}
\end{figure*}

Due to difficulties in the phase calibration, we were only able to produce a high quality image of J0322--483 at 2.4\,GHz, shown in Figure~\ref{fig:j0322-482}. We do not resolve PMN\,J0322--4820 and it is confined to the size of the beam: $56\times40$\,mas. The final image was made using a robust parameter of $+0.5$, and by flagging the Hartebeesthoek antenna, thus the beam size for PMN\,0322--4820 compared to MRC\,0225--065 for the same frequency is much larger. Details of the image properties are presented in Table~\ref{tab:lba_image_properties}. Compared to the spectral model fit to the ATCA and 2014 MWA observations, 18\% of the flux density was missing. We used a reported photometric redshift for PMN\,J0322--4820 of 0.16 \citep{2014ApJS..210....9B}, thus 1\,mas corresponds to a linear size of 2.650\,pc. We place an upper limit on the source size of 148\,pc. 

\section{Discussion}
\label{sec:discussion}
In this section, we will present a comprehensive analysis of both MRC\,0225--065 and PMN\,J0322--4820 to produce a unified perspective of these two sources with the aim of concluding whether they are young or frustrated PS sources. In Section~\ref{subsec:ls_vs_t}, we present our two sources in the linear size and turnover relation, in Section~\ref{subsec:hostgals}, we discuss the host galaxy properties according to mid-infrared, optical observations and radio properties.

\subsection{Linear Size and Turnover Relation}
\label{subsec:ls_vs_t}
\begin{figure}[htpb]
    \centering
    \includegraphics[width=0.9\textwidth]{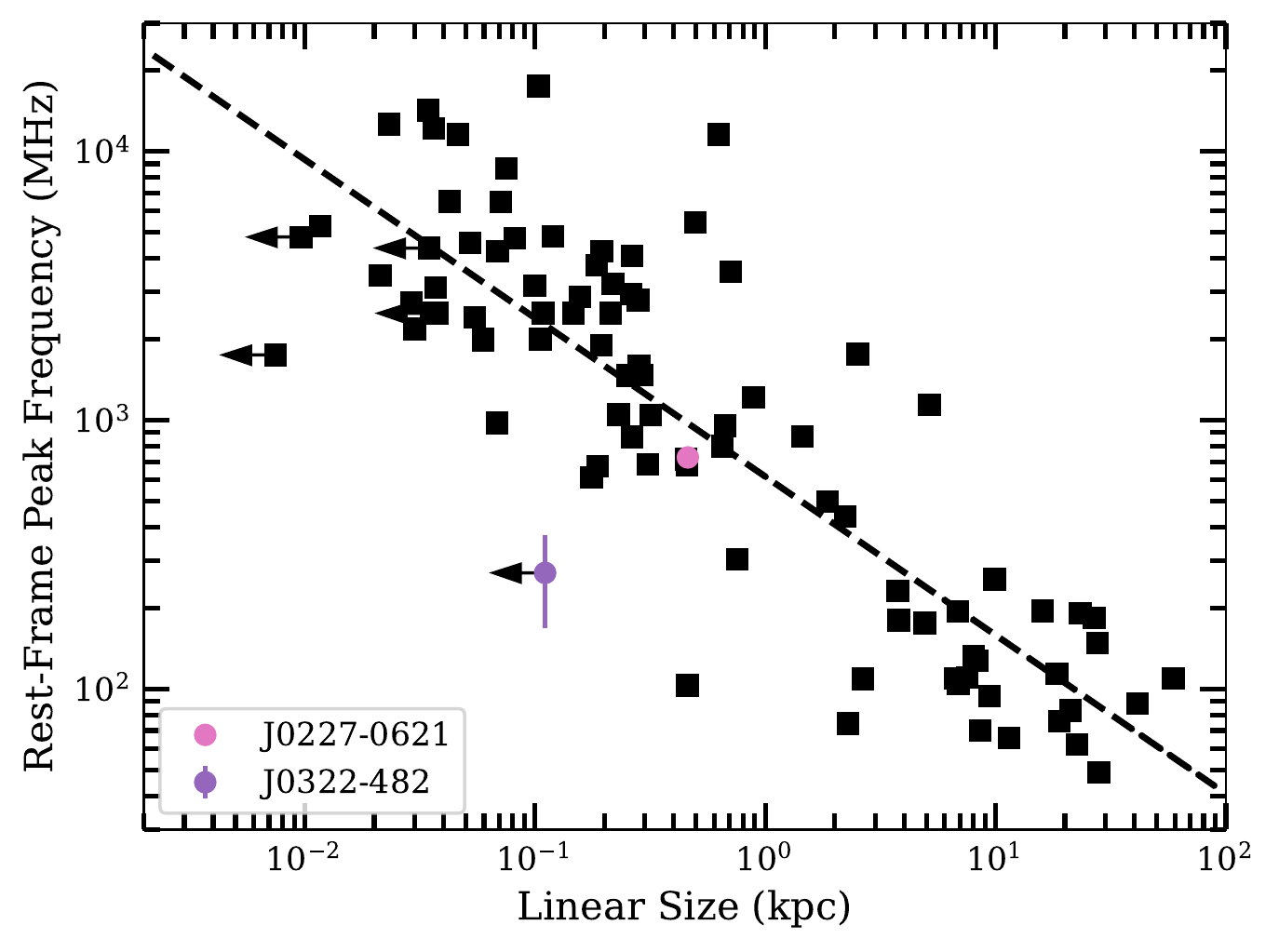}
    \caption[Linear Size vs Intrinsic Peak Frequency]{Rest frame peak frequency versus linear size. Sources in black are described in \citet{2019A&A...628A..56K}. The dashed line is the fit to the relation found by \citet{2014MNRAS.438..463O}. Arrows indicate maximum linear sizes for unresolved sources. MRC\,0225--065 (pink circle) and PMN\,J0322--4820 (purple circle) are plotted with linear sizes calculated from LBA images. The error bars for MRC\,0225--065 represent the range for peak frequencies calculated in R21. }
     \label{fig:ls_vp_j0227-0621}
\end{figure}

PS sources follow an inverse relation between their linear size and intrinsic turnover frequency, often referred to as the linear size turnover relation, first presented by \citet{1998PASP..110..493O}. This relation is directly predicted from the youth scenario \citep{1998PASP..110..493O} where the peak frequency is due to SSA and thus the linear size is directly related to the peak frequency \citep[][]{1981ARA&A..19..373K}. While modifications to models in the frustration scenario can reproduce this relation \citep{2018MNRAS.475.3493B}, it is generally understood that PS sources that fall below the linear size-turnover relation are likely compact beyond what is expected for a young source and a thus assumed to be frustrated. We plot both MRC\,0225--065 and PMN\,J0322--4820 on the linear size-turnover relation in Figure~\ref{fig:ls_vp_j0227-0621}, along with other known PS sources, details of which are discussed by \citet{2019A&A...628A..56K}. It is evident from Figure~\ref{fig:ls_vp_j0227-0621}, that MRC\,0225--065 is entirely consistent with the relation whereas PMN\,J0322--4820 sits somewhat below the relation, particularly since the linear size is an upper limit. This would suggest MRC\,0225--065 is consistent with the youth scenario whereas PMN\,J0322--4820 may be frustrated. However, it is worth nothing, R21 identified PMN\,J0322--4820 as a variable PS source with a changing spectral shape, and thus concluded it was likely a blazar. Furthermore, R21 found the peak frequency changed from $\sim$320\,MHz in 2013 to $\sim$145\,MHz in 2014. As the peak frequency is variable and PMN\,J0322--4820 is known to exhibit a changing spectral shape, its position on the linear size-turnover relation will also vary, shown by the error bar in Figure~\ref{fig:ls_vp_j0227-0621} corresponding to the range of the peak frequency from 2013 to 2014. Most likely, PMN\,J0322--4820 is only a temporary PS source and thus should not be included in this relation nor when considering the PS population at large. 

\subsection{Host Galaxy Properties}
\label{subsec:hostgals}

\subsubsection{\textit{WISE} Colours}
\label{subsubsec:wise}
MIR colour selection techniques using the \textit{Wide-Field Infrared Survey Explorer} \citep[WISE]{2010AJ....140.1868W} are widely used to efficiently distinguish between AGN and star-forming galaxies. 

\textit{WISE} is a MIR all sky survey covering four photometric bands: 3.4, 4.6, 12, and 22\,$\mu$m referred to as W1, W2, W3, and W4 respectively. The MIR wavelengths are sensitive to the emission from hot dust in the torus of the AGN, allowing for the identification of AGN where X-ray and optical emission may be blocked by intervening gas and dust. This also makes AGN stand out from star-bursting galaxies or stars due to their extremely red MIR emission \citep{2015ApJ...813...45L}. Obscured AGN with red MIR emission have been identified by their MIR colours, often by their place in a colour-colour diagram \citep{2011ApJ...735..112J,2015ApJ...813...45L}. 
The bulk of sources centred around $W1-W2=1.2$ and $W2-W3=3$ correspond to the region typically associated with quasars and AGN. MRC\,0225--065 is found in the region typically associated with emission from star formation or stellar emission; i.e. there is no evidence of hot AGN dust, however, there is evidence for moderate star formation. As we know MRC\,0225--065 is an AGN, it is likely the emission at MIR is a combination of these two processes.


PMN\,J0322--4820 is well within the elliptical regime, thus has low emission from star formation and no evidence of hot AGN dust. Blazars are typically found to dominate the top right region of the \textit{WISE} colour-colour plot as the MIR emission is dominated by the emission of the blazar over the galaxy (and associated stellar emission). A compact morphology and variable spectral shape suggest PMN\,J0322--4820 is a blazar. However, the \textit{WISE} colours of PMN\,J0322--4820 suggest that the host galaxy is an elliptical with predominantly red optical emission. Therefore, the emission from the potential radio blazar is not dominant in the MIR. While it is more common to find blazars in the top right region of the \textit{WISE} colour-colour plot, the MIR colours, which suggest the host galaxy for PMN\,J0322--4820 is an elliptical, are still consistent with a blazar classification \citep{2015MNRAS.449.3191Y,2019ApJS..242....4D}.


\subsubsection{Optical Spectra}
\label{subsubsec:opticalspectra}
MRC\,0225--065 has an optical spectrum from the $13^\mathrm{th}$ data release of the Sloan Digital Sky Survey \citep[SDSS]{2017ApJS..233...25A}. From the fitted spectrum, \citet{2017ApJS..233...25A} report a spectroscopic redshift for MRC\,0225--065 of $z=0.445$ and classify it as a broad-line, starburst quasar. The spectrum additionally has low-ionisation nuclear emission-line region (LINER) properties, evident from the strong NII, SiII and OI lines. A LINER has a high energy radiation field. There is still debate about whether this is AGN emission or star formation, but likely the combination of the broad lines, strong OIII emission and radio-loudness of MRC\,0225--065 is evidence of AGN. From the broad H$\alpha$, we can calculate the velocity dispersion according to: 
\begin{equation}
    d(\mathrm{velocity}) = c \frac{d(\lambda)}{\lambda_0},
\end{equation}
where $c$ is the speed of light, $d(\lambda)$ is the wavelength dispersion from the spectral fit, and $\lambda_0$ is the rest-frame wavelength of H$\alpha$. Using the reported fit to the broad H$\alpha$ from SDSS where $\lambda_\mathrm{observed}=9486$\,\AA, we use the equivalent width, EW$=30\pm4$\,\AA, and find the velocity dispersion to be $900\pm100$\,km/s. This large velocity dispersion may be from an extreme star formation wind but it is also indicative of the broad-line regions from an AGN, which is more consistent given our radio observations identify MRC\,0225--065 as an AGN. The broad H$\alpha$, and large velocity dispersion, is consistent with an AGN that is quite obscured, as reported by \citet{2017ApJS..233...25A} who classify it as a broad-line quasar. Perhaps of more interest are the starburst properties of MRC\,0225--065, namely OII and OIII emission lines, identified by \citet{2017ApJS..233...25A}. Both OII and OIII are forbidden lines with different origins: OII is mostly due to star formation and thus is often used as an indicator for star formation in galaxies; OIII is due to an AGN and can be used as a proxy for the AGN bolometric luminosity. This is also consistent with the \textit{WISE} colours discussed in Section~\ref{subsubsec:wise}, which find MRC\,0225--065 consistent with a galaxy with emission coming from both the AGN and star formation. Combining the radio, MIR and optical properties of MRC\,0225--065, it is likely this galaxy has moderate star formation with an obscured AGN.

\subsection{Radio Properties of MRC B0225--065}
\label{subsec:radioproperties}
Combining the spectral information and high resolution resolved structure of MRC\,0225--065, we are able to determine several intrinsic properties that can help differentiate between SSA and FFA models. In this section, we estimate the magnetic field strength and spectral ages to assess whether MRC\,0225--065 is consistent with the youth scenario. We do not consider PMN\,J0322--4820 in this section due to its unresolved morphology (even on mas scales) and since the radio variability suggests it is a blazar with an added beaming effect producing Doppler boosting and thus many of the assumptions required for these calculations no longer hold. 

\subsubsection{Magnetic Field}
\label{subsubsec:mag_fields}
As a means of evaluating the validity of SSA compared to an FFA, we can calculate the magnetic field estimates based on a pure SSA model and on equipartition. Equipartition assumes there is equal energy between the radiating particles and the magnetic field. The comparison between magnetic field estimates based on an SSA model and equipartition has been used as evidence both for the SSA model \citep[when the estimates are in agreement;][]{2008A&A...487..885O} and against \citep[when there is a clear disparity;][]{2019A&A...628A..56K}. In this section, we will first estimate the magnetic field assuming a purely SSA model, then assuming equipartition and compare these to determine whether SSA is a reasonable model for MRC\,0225--065. 

We can estimate the magnetic field strength, in Gauss, based on a purely SSA spectral model, $B_\mathrm{SSA}$, according to: 
\begin{equation}
    B_\mathrm{SSA} \approx \frac{(\nu_\mathrm{peak}/f(\alpha_\mathrm{thin}))^{5} {\theta_\mathrm{src, min}}^{2}{\theta_\mathrm{src,max}}^{2}}{{S_\mathrm{peak}}^2(1+z)},
    \label{eq:B_ssa}
\end{equation}
where $\nu_\mathrm{peak}$ is the observed peak frequency in GHz, $S_\mathrm{peak}$ is the flux density in Jy at the peak frequency for the source at redshift $z$ with angular minor and major component axis, $\theta_\mathrm{src,min}$ and $\theta_\mathrm{src,max}$, in mas \citep{1981ARA&A..19..373K}. We note, $f(\alpha_\mathrm{thin})$ is as defined by \citet{1981ARA&A..19..373K}, where it is loosely related to $\alpha_\mathrm{thin}$. We take $f(\alpha_\mathrm{thin})=8$ based on values from \citet{1983ApJ...264..296M,2008A&A...487..885O}.
 
Now, assuming equipartition, we calculate the magnetic field strength, in Gauss, according to \citep{1980ARA&A..18..165M}, as $B_\mathrm{equi}$ by assuming the component has cylindrical symmetry such that the width of the source on the sky is equivalent to the line of sight path-length.

For both calculations, we calculate $B_\mathrm{SSA}$ and $B_\mathrm{equi}$ for the compact core region rather than the total source, to ensure we are comparing a homogeneous region \citep{2008A&A...487..885O,2019A&A...628A..56K}. For MRC\,0225--065, using Equation~\ref{eq:B_ssa}, we estimate the magnetic field strength for a purely SSA model to be $B_\mathrm{SSA}\approx$6$\pm$7\,mG for the core region where $\theta_\mathrm{src}=2.5\times 4$\,mas. To estimate $B_\mathrm{equi}$, we assume a filling factor $\eta=1$ and set $k=1$\footnote{$k=1$ is equivalent to the minimum energy condition, however values for $k$ have ranged from 1 to 100, where $k=100$ produces an order of magnitude difference in $B_\mathrm{equi}$ \citep{1971ranp.book.....P,1980ARA&A..18..165M}} and find $B_\mathrm{equi}\approx$6$\pm$2\,mG. As $B_\mathrm{SSA}$ is within the uncertainties of $B_\mathrm{equi}$, it suggests the core region of MRC\,0225--065 is in equipartition and consistent with a pure SSA model. While this does not exclude the FFA model, it does provide supportive evidence for the SSA model. Furthermore, it may not be a valid assumption that MRC\,0225--065 is in equipartition, thus the equation from \citet{1980ARA&A..18..165M} for $B_\mathrm{equi}$ would not be a reasonable estimate of the magnetic field strength. 

We can also use the estimated magnetic field to calculate the age of the electron population as a proxy for the age of the jets/lobes. Calculating the spectral age of the electron population requires an accurate estimate of the break frequency, $\nu_b$. We can thus calculate the spectral age, $\tau_\mathrm{spec}$, according to: 
\begin{equation}
\begin{split}
    \tau_\mathrm{spec} &= \frac{aB^{1/2}}{B^2+{B_\mathrm{iC}}^{2}}\left[\nu_b (1+z) \right]^{-1/2} \\
    \mathrm{where} \\
    B_\mathrm{iC} &= 0.318(1+z)^{2} \\
    a &= \left(\frac{243\pi {m_e}^{5}c^2}{4{\mu_0}^2e^7} \right)^{1/2}
\end{split}
    \label{eq:spectral_age}
\end{equation}
where $B_\mathrm{iC}$ is the magnitude of the microwave background magnetic field in nT, $B$ is the magnetic field of the source in nT, $\nu_b$ is the break frequency in GHz, and the constants $m_e$, $c$, $\mu_0$, and $e$ are the mass of an electron, speed of light, magnetic permeability of free space, and charge of an electron, respectively. 

It is possible the core is actually an unresolved double of more recent AGN activity than the outer lobes, producing the steep ($\alpha\lesssim-1$, see Table~\ref{tab:j0227-0621_lba_components}) spectral index. We assume a constant expansion speed, $v$, and use the linear sizes to estimate the dynamical age, $\tau_\mathrm{dyn}$, of the core and outer lobes. Using the magnetic field calculated for the core region assuming equipartition, i.e. setting $B=B_\mathrm{equi}=6\pm2$\,mG, and determining a break frequency, we can estimate the spectral age of the core. Using a break frequency of $\nu_b=14.3\pm2.7$\,GHz, calculated from the double SSA spectral model fit, we estimate the spectral age of the core to be $\tau_\mathrm{spec}\approx700\pm100$\,years. We then calculate an upper limit on the expected expansion velocity of $v\leq0.13$\,$c$ (using simple speed = distance/time arguments) for the core using the upper limit for the linear source size of $\theta_\mathrm{src}\leq26$\,pc, as outlined in Section~\ref{sec:images_j0227-0621}. An expansion velocity of $v=0.13$\,$c$ is well within previous measurements of the expansion speeds for compact AGN that have been found to range from 0.1\,$c$ up to 0.7\,$c$ \citep{2003PASA...20...69P,2012ApJ...760...77A,2020MNRAS.499.1340O}. The range of expansion velocities would correspond to a range in dynamical ages for the core of $100\lesssim \tau_\mathrm{dyn} \lesssim 900$\,years. If we assume the expansion velocity of the core of ``inner lobes" is roughly equal to that of the outer lobes from a previous epoch of activity, we can place an upper limit on the dynamical ages of the outer lobes. We calculate the distance between the core and L1 as $\sim210$\,pc, which corresponds to a dynamical age of 5000\,years for an expansion velocity of $0.13$\,$c$. For the range of dynamical ages for typical PS sources, we expect the age of the outer lobes to be $1000\lesssim \tau_\mathrm{dyn} \lesssim 7000$\,years. Previous estimates for the ages of PS sources using similar assumptions have estimated ages from $\sim10^1$ to $\sim10^5$\,years \citep{2010MNRAS.402.1892O}, which is entirely consistent with our age estimates for both the inner core and outer lobes. 

As the ages, expansion velocities, and magnetic fields that we calculate are all consistent with the SSA model and a youth scenario, it appears MRC\,0225--065 is more consistent with a young CSO rather than a frustrated compact AGN. However, there are several caveats and assumptions made in these calculations. Thus, while these results are consistent with the evolutionary scenario of MRC\,0225--065 being the youth model, it is not sufficient for excluding the frustration scenario entirely.

\section{A Unified Perspective of MRC B0225--065 and PMN J0322--4820}
\label{sec:unified_perspective}
Combining all the information we have obtained about MRC\,0225--065, we begin to create a unified perspective that suggests MRC\,0225--065 is a CSO with a peaked spectrum best explained by SSA and recent jet activity over the last $10^2$--$10^3$\,years. A summary of the evidence in support of this conclusion are as follows: 
\begin{itemize}
    \item \textbf{Variability: }R21 identified spectral variability of MRC\,0225--065 with a constant spectral shape, consistent with variability due to RISS. Further spectral variability monitoring by R22 detected no further variability, suggesting a resolved structure but consistent PS source classification. This observation suggests it is unlikely MRC\,0225--065 is a contaminating blazar or source with only a temporary PS source classification, such as frustrated sources with an inhomogeneous surrounding medium. 
    \item \textbf{Radio morphology: }Previously, it has been suggested frustrated PS sources are more likely to show an asymmetrical morphology due to the asymmetrical environment confining the growth of the lobes. Inversely, this suggests young PS sources that are not frustrated may be more likely to show a symmetrical morphology like that of a CSO. MRC\,0225--065 has a very symmetrical morphology according to our LBA images, suggesting it may not be interacting with its surrounding environment. 
    \item \textbf{Linear size and turnover relation: }We find MRC\,0225--065 is entirely consistent with the linear size turnover relation, a natural product of the youth scenario. Although, it can be reproduced in certain frustration models. 
    \item \textbf{Host galaxy: }Using the MIR colours reported in by \textit{WISE} and the optical spectrum from SDSS, we identify the MRC\,0225--065 as having an obscured AGN with moderate star formation. Since the AGN does not dominate the entire MIR and optical emission, and there is still star formation present, it is possible the AGN has only recently been switched on and thus has not yet quenched all star formation in the galaxy, which is not surprising given the compact size of MRC\,0225--065. 
    \item \textbf{Magnetic field: }Estimating the magnetic field using a purely SSA model and comparing it to the magnetic field calculated assuming equipartition are entirely consistent, suggesting the SSA model is a reasonable model for MRC\,0225--065
    \item \textbf{Spectral ages: }Using spectral modelling of the break frequency, we estimate the age of the radio emission (from the core and lobes) to be roughly 700\,years, consistent with estimates of the age of PS sources in the youth scenario. 
    \item \textbf{Dynamical ages: }Using the linear size from our LBA images and previous measurements of expansion velocity we estimate MRC\,0225--065 has two major epochs of activity, one between 1000 to 7000\,years ago and another more recently from 100 to 900\,years ago. This is also consistent with previous estimates of the ages for young PS sources. Furthermore, due to the missing flux density at 8.3\,GHz, this estimate should be considered an upper limit as the spectral indices for each component may be artificially steepened by the missing flux density. 
\end{itemize}
We therefore conclude, MRC\,0225--065 is likely a young AGN and with the peak occurring due to SSA. 

Likewise, combining all information of PMN\,J0322--4820, we can also begin to create a unified picture that PMN\,J0322--4820 is a blazar. A summary of the evidence for this conclusion are: 
\begin{itemize}
    \item \textbf{Spectral variability: }R21 identified PMN\,J0322--4820 as a variable source in and classified it as showing a changing spectral shape. The dramatic change in spectral shape in the megahertz regime on a timescale of $\sim1$\,year is inconsistent with evolutionary models for PS sources and predicted variability due to RISS. The changing spectral shape is most easily explained by the dynamical nature of blazars. 
    \item \textbf{Radio morphology: }The high resolution image of PMN\,J0322--4820 using the LBA found it was still compact on mas scales. This is also entirely consistent with a blazar morphology, which appears compact due to orientation effects. 
    \item \textbf{Linear size and turnover relation: }PMN\,J0322--4820 sits well below the linear size and turnover relation typically associated with PS sources. This could either be because it is a frustrated source and is thus more compact than expected for it's predicted age. However, more likely, is that the temporary peak detected with the MWA in 2014 was a result of the variability of a blazar with effects like Doppler boosting influencing measurements and thus the spectral peak is unrelated to the source age or absorption mechanisms. 
    \item \textbf{WISE MIR Colours: }PMN\,J0322--4820 has \textit{WISE} colours typically associated with elliptical galaxies and/or LERGs/BL Lac blazars.  
\end{itemize}
We therefore identify PMN\,J0322--4820 as a new blazar where the jets are oriented along the line-of-sight. However, PMN\,J0322--4820 was not in the ROMA-bzcat catalogue of $\gamma$-ray emitting blazars. This is potentially due to the steep spectrum at frequencies over 1\,GHz where PMN\,J0322--4820 is too faint to be detected by traditional blazar searches. We suggest further observations using higher frequency observations in the X-ray or $\gamma$ regimes to search for any high frequency counterpart \citep{2009A&A...495..691M,2015Ap&SS.357...75M}. We conclude PMN\,J0322--4820 should not be included in any future population studies of PS sources as it is a contaminating blazar and not a genuine PS source. Furthermore, this highlights the possibility of a population of blazars with steep spectra at high frequencies ($\nu\geq1$\,GHz) that aren't detected in traditional blazar searches and thus may be contaminating populations of PS sources. Low-frequency spectral variability thus presents as a new method for identifying blazar candidates.

\section{Conclusion}
\label{sec:conc}
We have sought to compare detections of spectral variability for two PS sources with small scale ($\sim$mas) morphology and structures. The images produced using observations with the LBA have identified one resolved and one unresolved PS source. We have also combined our observations with archival observations of the host galaxies of our sources to provide evidence for either the youth or frustration scenario.

We find PMN\,J0322--4820 is unresolved with the LBA at 2.4\,GHz, and pace an upper limit of the source size to be 148\,pc, using a photometric redshift of 0.16. In R21, PMN\,J0322--4820 was found to show a changing spectral shape and was presented as a blazar candidate. Comparing our compact morphology with the spectral variability of R21, we find PMN\,J0322--4820 is consistent with a blazar classification, and suggest high frequency (X-ray or Gamma) to confirm.


We resolve MRC\,0225--065 into three components at both 2.4\,GHz and 8.3\,GHz: a bright central region containing $\sim$50\% of the total flux density, and two fainter regions roughly equal distance from the central region. In R21 and R22, MRC\,0225--065 was found to show low levels of variability with a constant spectral shape, and presented as showing variability due to ISS from a compact morphology with resolved structure on mas scales. We find the projected linear size to be 430\,pc, using a spectroscopic redshift of 0.445. Using spectral modelling, we calculate the magnetic field assuming a purely SSA model, and find it is in agreement with the magnetic field calculated assuming equipartition. We therefore conclude MRC\,0225--065 is a young CSO, with a PS classification due to SSA. We found the core to have a spectral age of $\tau_\mathrm{spec}=700\pm100$\,years, which is consistent with previous age estimates of young CSO sources of $10^1$ -- $10^5$\,years \citep{2010MNRAS.402.1892O,2020MNRAS.499.1340O}. Furthermore, we use the spectral age of the core and the upper limit of core size to calculate and expected expansion velocity (assuming the simple relation $\mathrm{speed}=\mathrm{distance/time}$), and place an upper limit on the expansion velocity of the lobes to be $v=0.13c$, well within previous measurements of expansion velocities for PS sources of $0.1c\lesssim v \lesssim 0.7c$ \citep{2020MNRAS.499.1340O}. Lastly, we use this to estimate the dynamical age of the outer lobes and estimate their age to be $\tau_\mathrm{dyn}\approx5000$\,years, again, well within previous estimates of ages for young PS sources. 

Our findings highlight the advantage of spectral variability in identifying different milliarcsecond structures in PS sources traditionally acquired using VLBI. Furthermore, we have confirmed the use of identifying contaminating sources displaying only a temporary spectral peak and present spectral variability as a new method for identifying steep spectrum blazars. We also suggest future observations of MRC\,0225--065 to search for direct observations of expansion to better constraining the expansion velocity and age. We recommend observations of MRC\,0225--065 with the VLBA for improved sensitivity and more $u,v$-coverage on short baselines to recover more flux density from extended structures. Likewise, with improved accuracy of the position for MRC\,2236-454, we suggest another VLBI observation. 

\begin{acknowledgement}
 We thank the referees for their comments that improved the overall quality of this work. KR acknowledges a Doctoral Scholarship and an Australian Government Research Training Programme scholarship administered through Curtin University of Western Australia. JRC thanks the Nederlandse Organisatie voor Wetenschappelijk Onderzoek (NWO) for support via the Talent Programme Veni grant. NHW is supported by an Australian Research Council Future Fellowship (project number FT190100231) funded by the Australian Government. The Long Baseline Array is part of the Australia Telescope National Facility \url{https://ror.org/05qajvd42} which is funded by the Australian Government for operation as a National Facility managed by CSIRO. This work was supported by resources provided by the Pawsey Supercomputing Centre with funding from the Australian Government and the Government of Western Australia. LBA data was correlated at the Pawsey Supercomputer Centre using the DiFX software \citep{2011PASP..123..275D}. This scientific work uses data obtained from Inyarrimanha Ilgari Bundara/the Murchison Radio-astronomy Observatory. We acknowledge the Wajarri Yamaji People as the Traditional Owners and native title holders of the Observatory site. The Australian SKA Pathfinder is part of the Australia Telescope National Facility \url{https://ror.org/05qajvd42} which is managed by CSIRO. Operation of ASKAP is funded by the Australian Government with support from the National Collaborative Research Infrastructure Strategy. ASKAP uses the resources of the Pawsey Supercomputing Centre. Establishment of ASKAP, the Murchison Radio-astronomy Observatory and the Pawsey Supercomputing Centre are initiatives of the Australian Government, with support from the Government of Western Australia and the Science and Industry Endowment Fund. This paper includes archived data obtained through the CSIRO ASKAP Science Data Archive, CASDA (https://data.csiro.au). This research made use of NASA's Astrophysics Data System, the VizieR catalog access tool, CDS, Strasbourg, France. We also make use of the \textsc{IPython} package \citep{Ipythoncite}; SciPy \citep{2020SciPy-NMeth};  \textsc{Matplotlib}, a \textsc{Python} library for publication quality graphics \citep{Hunter:2007}; \textsc{Astropy}, a community-developed core \textsc{Python} package for astronomy \citep{astropy:2013, astropy:2018}; \textsc{pandas}, a data analysis and manipulation \textsc{Python} module \citep{reback2020pandas,mckinney-proc-scipy-2010}; and \textsc{NumPy} \citep{vaderwalt_numpy_2011}. We also made extensive use of the visualisation and analysis packages DS9\footnote{\href{ds9.si.edu}{http://ds9.si.edu/site/Home.html}} and Topcat \citep{2005ASPC..347...29T}. This work was compiled in the useful online \LaTeX{} editor Overleaf.
\end{acknowledgement}

\bibliography{bib}

\appendix

\end{document}